\documentstyle[twocolumn,aps,amsfonts]{revtex}
\def\dx{d^{2}x}
\def\dy{d^{2}y}
\def\B{{\cal B}}
\def\upA{\uparrow}
\def\dnA{\downarrow}
\def\a{{\cal A}}
\def\O{{\cal O}}
\def\P{{\cal P}}
\def\Q{\widehat{\O}}
\def\bpmatrix{\left(\begin{array}{c}}
\def\epmatrix{\end{array}\right)}

\def\F{{\frak S}}
\def\FF{\widehat{\F}}
\def\Laugh{\F_{\text{LN}}[\bbox{x}]}
\begin{document}
\title{Improved Composite-Boson Theory of Quantum Hall Ferromagnets\\
and Skyrmions without Sigma Models}
\author{Z.F. Ezawa}
\address{Department of Physics, Tohoku University, Sendai 980-8578, Japan}
\maketitle
\begin{abstract}
We propose an improved composite-boson theory of quantum Hall ferromagnets, 
where the field operator describes solely the physical degrees of freedom 
representing the deviation from the ground state.  In this scheme skyrmions 
appear merely as generic excitations confined in the lowest Landau level.  We 
evaluate the excitation energy of one skyrmion.  Our theoretical estimation 
accounts for the activation-energy data due to Schmeller et al. remarkably 
well.
\end{abstract}

{\par{\bf Introduction}:\ }
Low-energy charged excitations are skyrmions \cite{Skyrmion} in quantum Hall (QH) 
ferromagnets when the Zeeman energy ($g^{*}\mu _{B}B$) is small.  Skyrmions were 
initially considered as solutions of the effective nonlinear sigma (NL$\sigma $) 
model \cite{SkyrmQH}, and later studied also in a Hartree-Fock approximation 
\cite{SkyrmHartFock}.  Their existence has been confirmed experimentally 
\cite{SkyExp,SkyExpEneA,SkyExpEneB}.  The skyrmion has an intrinsic scale $\kappa $, 
which is a function of the gyromagnetic factor $g^{*}$ ($\kappa \rightarrow 0$ as $g^{*}\rightarrow \infty $ and $\kappa \rightarrow \infty $ 
as $g^{*}\rightarrow 0$).  

However, the NL$\sigma $-model approach suffers from problems.  Basically, 
the NL$\sigma $ model describes correctly perturbative spin fluctuations 
\cite{EzaIQC,MG,KallinHalperin}, but there is no reason a priori that it is 
capable to treat nonperturbative objects such as skyrmions.  We mention 
explicitly three points at issue.  First, it is an unsatisfactory theory since 
the relation between the microscopic wave function and the classical solution 
of skyrmion is totally unclear.  Second, it is an inconsistent theory because 
the skyrmion and vortex excitations are treated entirely different objects 
though their wave functions become identical as $\kappa \rightarrow 0$.  Third, the predicted 
skyrmion energy \cite{SkyrmQH,SkyrmA} is too large compared with experimental data 
\cite{SkyExpEneA,SkyExpEneB}.  In particular, the activation energy $\Delta $ has been 
predicted that $\Delta =\sqrt {\pi /8}E_{C}^{0}$ as $g^{*}\rightarrow 0$, with $E_{C}^{0}=e^{2}/\varepsilon \ell _{B}$ the Coulomb energy 
unit.  On the contrary, the observed value is consistent with $\Delta \simeq 0$ as 
revealed in recent experiments \cite{SkyExpEneB}.

In this paper we propose a new scheme of skyrmions based on an improved 
composite-boson (CB) theory \cite{EzaICBa}.  In this scheme the field operator 
describes solely the physical degrees of freedom representing the deviation 
from the ground state, and the semiclassical property of excitations is 
determined directly by their microscopic wave functions.  Charged excitations 
are skyrmions and vortices, which appear merely as generic topological 
excitations confined within the lowest Landau level (LLL).  Their excitation 
modulates the electron density $\rho (\bbox{x})$ according to the soliton equation 
\cite{EHIc,EzaIQC},
\begin{equation}
{\nu \over 4\pi }\bbox{\nabla }^{2}\ln \rho (\bbox{x}) - \rho (\bbox{x}) + \rho _{0}= \nu Q(\bbox{x}) ,
\label{SolitEq}
\end{equation}
where $\nu $ is the filling factor, $\rho _{0}$ is the average density and $Q(\bbox{x})$ is the 
topological charge density determined by the wave function.  The main part of 
the excitation energy is the Coulomb energy induced by this density 
modulation.  The skyrmion excitation energy is shown to vanish as $g^{*}\rightarrow 0$.  Our 
formalism is free from the difficulties mentioned above.  Throughout the paper 
we use the natural units $\hbar =c=1$.
{\par{\bf Vortices}:\ }
Leaving the derivation of the improved CB theory later, we summarize first its 
idea applied to the single-component QH state.  We denote the electron field 
by $\psi (\bbox{x})$ and its position by the complex coordinate normalized as 
$z=(x+iy)/2\ell _{B}$.  Any state $|\F\rangle $ at $\nu =1/m$ ($m$ odd) is represented by the 
wave function,
\begin{equation}
\F[\bbox{x}] \equiv  \langle 0|\psi (\bbox{x}_{1})\cdots \psi (\bbox{x}_{N})|\F\rangle  = \omega [z]\Laugh ,
\label{WaveElect}
\end{equation}
where $\Laugh=\prod _{r<s}(z_{r}-z_{s})^{m}\exp[-\sum _{r=1}^{N}|z_{r}|^{2}]$ is the Laughlin function, and 
$\omega [z]\equiv \omega (z_{1},z_{2},\cdots ,z_{N})$ is an analytic function symmetric in all $N$ variables.  
The mapping from the fermionic wave function $\F[\bbox{x}]$ to the bosonic function 
$\omega [z]$ defines a bosonization.  We call the underlying boson the 
\textit{dressed composite boson} and denote its field operator by $\varphi (\bbox{x})$.  We 
derive that
\begin{equation}
\F_{\varphi }[\bbox{x}] \equiv  \langle 0|\varphi (\bbox{x}_{1})\cdots \varphi (\bbox{x}_{N})|\F\rangle  = \omega [z].
\label{WaveFunctDress}
\end{equation}
The Laughlin state is represented by $\F_{\varphi }[\bbox{x}]=1$.  A typical vortex state is 
by $\F_{\varphi }[\bbox{x}]=\prod _{r}^{N}z_{r}$, implying that $\langle \varphi (\bbox{x})\rangle =z$ in the semiclassical 
approximation.  This is a highly nontrivial constraint leading to the soliton 
equation (\ref{SolitEq}), where $Q(\bbox{x})\equiv Q^{V}(\bbox{x})$ is the vorticity density,
\begin{equation}
Q^{V}(\bbox{x})={1\over 2\pi i}\varepsilon _{ij}\partial _{i}\partial _{j}\ln\langle \varphi (\bbox{x})\rangle  =\delta (\bbox{x}).
\end{equation}
The vortex carries the electron number,
\begin{equation}
\Delta N=\int d^{2}x[\rho (\bbox{x})-\rho _{0}]=-\nu , 
\label{ElectNumbe}
\end{equation}
as follows from the soliton equation.
{\par{\bf Skyrmions}:\ }
We next summarize the idea of an improved CB theory applied to QH 
ferromagnets.  We denote the spin component by the index $\alpha (=\upA ,\dnA )$.  Any state 
at $\nu =1/m$ is represented by the wave function similar to (\ref{WaveElect}).  Let 
us explicitly consider the case when the spinor component is factorizable,
\begin{equation}
\F[\bbox{x}] = \prod _{r}\bpmatrix \omega ^{\upA }(z_{r})\\ \omega ^{\dnA }(z_{r})\epmatrix_{\kern-1pt r} \Laugh.
\label{SkyrmWave}
\end{equation}
The ground state is given by $\omega ^{\upA }(z)=1$ and $\omega ^{\dnA }(z)=0$.  From a set of two 
analytic functions $\omega ^{\alpha }(z)$ we construct the complex-projective (CP) field 
$\bbox{n}(\bbox{x})$ whose two components $n^{\alpha }(\bbox{x})$ are
\begin{equation}
n^{\alpha }(\bbox{x}) = {\omega ^{\alpha }(z)\over \sqrt {|\omega ^{\upA }(z)|^{2}+|\omega ^{\dnA }(z)|^{2}}}. 
\label{GenerSkyrm}
\end{equation}
This represents the most general skyrmion configuration \cite{Skyrmion}.  The 
normalized spin field, or the nonlinear sigma field, is defined by 
$s^{a}(\bbox{x})=\bbox{n}(\bbox{x})^{\dagger }\tau ^{a}\bbox{n}(\bbox{x})$ with $\tau ^{a}$ the Pauli matrices.  The skyrmion configuration 
is indexed by the Pontryagin number \cite{Skyrmion}, whose density is
\begin{equation}
Q^{P}(\bbox{x}) = {1\over 8\pi }\varepsilon _{abc}\varepsilon _{ij}s^{a}(\bbox{x})\partial ^{i}s^{b}(\bbox{x})\partial ^{j}s^{c}(\bbox{x}) .
\label{PontrNumbe}
\end{equation}
The skyrmion excitation modulates not only the spin density but also the 
electron density according to the soliton equation (\ref{SolitEq}), where 
$Q(\bbox{x})\equiv Q^{P}(\bbox{x})$ is the Pontryagin number density (\ref{PontrNumbe}).  

The simplest excitation is given by one skyrmion with scale $\kappa $ sitting 
at $\bbox{x}=0$, whose wave function is \cite{MG}.
\begin{equation}
\F[\bbox{x}] = \prod _{r}\bpmatrix z_{r}\\ \kappa /2 \epmatrix_{\kern-1pt r} \Laugh.
\label{SkyrmWaveS}
\end{equation}
In the limit $\kappa \rightarrow 0$ the skyrmion is reduced to the vortex created in the 
spin-polarized ground state.  The scale $\kappa $ is to be fixed dynamically to 
minimize the excitation energy.  The Pontryagin number density (\ref{PontrNumbe}) 
is calculated for the simplest skyrmion (\ref{SkyrmWaveS}) as
\begin{equation}
Q^{P}(\bbox{x})= {1\over \pi } {(\kappa \ell _{B})^{2}\over [r^{2}+(\kappa \ell _{B})^{2}]^{2}} .
\label{SolitEqSkyrmB}
\end{equation}
The skyrmion carries the electron number (\ref{ElectNumbe}), as follows from the 
soliton equation (\ref{SolitEq}) with $Q(\bbox{x})=Q^{P}(\bbox{x})$.
{\par{\bf Bosonization}:\ }  We now construct the improved CB theory 
explicitly.  We start with the kinetic Hamiltonian for planar electrons in 
external magnetic field 
$(0,0,-B)$,
\begin{equation}
H_{K}= {1\over 2M}\int \dx \psi ^{\dagger }(\bbox{x})(P_{x}-iP_{y})(P_{x}+iP_{y})\psi (\bbox{x}) ,
\label{HamilB}
\end{equation}
where $P_{j}=-i\partial _{j}+eA_{j}$ is the covariant momentum with $A_{j}={1\over 2}\varepsilon _{jk}x_{k}B$.  Here, 
$\varepsilon _{12}=-\varepsilon _{21}=1$ and $\varepsilon _{11}=\varepsilon _{22}=0$.  

We define the \textit{bare} CB field by way of an operator phase 
transformation of the electron field $\psi (\bbox{x})$, $\phi (\bbox{x})=e^{-i\Theta (\bbox{x})}\psi (\bbox{x})$.  The phase 
field $\Theta (\bbox{x})$ is chosen to attach $m$ units of Dirac flux quanta $2\pi /e$ to each 
electron via the relation, $\varepsilon _{ij}\partial _{i}\partial _{j}\Theta (\bbox{x})=2\pi m\rho (\bbox{x})$, where $\rho (\bbox{x})\equiv \psi ^{\dagger }(\bbox{x})\psi (\bbox{x})$ is 
the electron density.  When $m$ is odd, $\phi (\bbox{x})$ is proved to be a bosonic 
operator.  The covariant momentum for the bare CB field is 
$\check{P}_{k}=P_{k}+\partial _{k}\Theta \equiv -i\partial _{k}-e\varepsilon _{kj}\partial _{j}\a(\bbox{x})$.  The bare CB feels the effective magnetic field 
$\B_{\text{eff}}=e^{-1}\bbox{\nabla }^{2}\a(\bbox{x})$, which vanishes on the homogeneous state $\langle \rho (\bbox{x})\rangle =\rho _{0}$ 
realized at $\nu =1/m$.  The auxiliary field $\a(\bbox{x})$ is solved as
\begin{equation}
\a(\bbox{x}) = m\int \dy \ln\biggl({|\bbox{x}-\bbox{y}|\over 2\ell _{B}}\biggr) \varrho  (\bbox{y}) ,
\label{SpinB}
\end{equation}
with $\varrho  (\bbox{y})\equiv \rho (\bbox{y})-\rho _{0}$.  The bare CB is the one used in literatures 
\cite{LGCSx,EzaIQC}.

We proceed to define the {\it dressed} CB field $\varphi (\bbox{x})$, 
\begin{equation}
\varphi (\bbox{x})=e^{-\a(\bbox{x})}\phi (\bbox{x}) = e^{-\a(x)-i\Theta (\bbox{x})}\psi (\bbox{x}) ,
\label{DressField}
\end{equation}
by dressing the bare CB with a cloud of the effective magnetic field described 
by $\a(\bbox{x})$.  Substituting (\ref{DressField}) into (\ref{HamilB}), the kinetic 
Hamiltonian is transformed into 
\begin{equation}
H_{K} = {1\over 2M}\int d^{2}x \varphi ^{\ddag }(\bbox{x})(\P_{x}-i\P_{y})(\P_{x}+i\P_{y})\varphi (\bbox{x}) ,
\label{HamilCB}
\end{equation}
where we have defined $\varphi ^{\ddag }(\bbox{x})\equiv \varphi ^{\dagger }(\bbox{x})e^{2\a(\bbox{x})}$, with which 
$\rho (\bbox{x})=\psi ^{\dagger }(\bbox{x})\psi (\bbox{x})=\varphi ^{\ddag }(\bbox{x})\varphi (\bbox{x})$, and 
\begin{equation}
\P_{j} =  -i\partial _{j} - (\varepsilon _{jk} + i\delta _{jk})\partial _{k}\a(\bbox{x}).
\label{CovarMomenR}
\end{equation}
Analyzing the Lagrangian density we find that the canonical conjugate of 
$\varphi (\bbox{x})$ is not $i\varphi ^{\dagger }(\bbox{x})$ but $i\varphi ^{\ddag }(\bbox{x})$.  A type of the field operator 
(\ref{DressField}) was first considered by Read\cite{ReadA} and revived recently by 
Rajaraman et al.\cite{RajaramanCB}.  

We suppress the kinetic energy by imposing the LLL condition,
\begin{equation}
(\P_{x}+i\P_{y})\varphi (\bbox{x})|\F\rangle =-{i\over \ell _{B}}{\partial \over \partial z^{*}}\varphi (\bbox{x})|\F\rangle  = 0 .
\label{LLLcondiDress}
\end{equation}
Solving this condition we find that the $N$-body wave function $\F_{\varphi }[\bbox{x}]$ is an 
analytic function as in (\ref{WaveFunctDress}).  It is an easy exercise to derive 
the following relation\cite{RajaramanCB},
\begin{eqnarray*}
\varphi ^{\ddag }(\bbox{x}_{1})\cdots \varphi ^{\ddag }(\bbox{x}_{N})|0\rangle  = \Laugh \psi ^{\dagger }(\bbox{x}_{1})\cdots \psi ^{\dagger }(\bbox{x}_{N})|0\rangle  .
\end{eqnarray*}
Because of this relation the function $\omega [z]$ in the wave function (\ref{WaveElect}) 
is given precisely by the formula (\ref{WaveFunctDress}).  

When the wave function is factorizable, $\F[z]=\prod _{r}\omega (z_{r})$, the one-point 
function is given by $\langle \varphi (\bbox{x})\rangle =\omega (z)$.  Using (\ref{DressField}) we may set
\begin{equation}
e^{\a(\bbox{x})}e^{i\chi (\bbox{x})}\sqrt {\rho (\bbox{x})} = \omega (z),
\label{preSolitMono}
\end{equation}
because $\rho (\bbox{x})=\phi (\bbox{x})^{\dagger }\phi (\bbox{x})$.  It is easy to see that the Cauchy-Rieman equation 
yields the soliton equation (\ref{SolitEq}).  Hence, the soliton equation may be 
viewed as a semiclassical LLL condition for topological excitations \cite{EHIc}.

One might question the hermiticity of the theory\cite{RajaramanCB}, since 
the covariant momentum (\ref{CovarMomenR}) has an unusual expression.  It is 
related with the fact that the canonical conjugate of $\varphi (\bbox{x})$ is 
$i\varphi ^{\ddag }(\bbox{x})\equiv i\varphi ^{\dagger }(\bbox{x})e^{2\a(\bbox{x})}$.  It implies that the hermiticity is defined together 
with the measure $e^{2\a(\bbox{x})}$.  Such a measure has arisen since the 
transformation (\ref{DressField}) is not unitary.  The covariant momentum $\P_{j}$ is 
hermitian together with this measure.  
{\par{\bf Quantum Hall Ferromagnet}:\ }
We proceed to analyze the QH system with the spin degree of freedom.  The 
electron field, the bare and dressed CB fields are denoted by $\psi ^{\alpha }(\bbox{x})$, $\phi ^{\alpha }(\bbox{x})$ 
and $\varphi ^{\alpha }(\bbox{x})$, respectively.  We also use the two-component electron field 
$\Psi (\bbox{x})$ and dressed CB field $\Phi (\bbox{x})$.  All formulas from (\ref{HamilB}) to 
(\ref{preSolitMono}) hold with appropriate modifications.  The phase field $\Theta (\bbox{x})$ 
and the auxiliary field $\a(\bbox{x})$ are defined by the same equations as before 
together with $\rho (\bbox{x})=\Psi ^{\dagger }(\bbox{x})\Psi (\bbox{x})=\Phi ^{\dagger }(\bbox{x})e^{2\a(\bbox{x})}\Phi (\bbox{x})$.  Now, the bare CB field is 
decomposed into the U(1) field $\phi (\bbox{x})$ and the SU(2) field $n^{\alpha }(\bbox{x})$, 
$\phi ^{\alpha }(\bbox{x})=\phi (\bbox{x})n^{\alpha }(\bbox{x})$ with $n^{\alpha }$ the CP field,
\begin{equation}
\varphi ^{\alpha }(\bbox{x}) = e^{-\a(\bbox{x})}\phi ^{\alpha }(\bbox{x})=e^{-\a(\bbox{x})}\phi (\bbox{x})n^{\alpha }(\bbox{x}) .
\label{BasicFormu}
\end{equation}
The spin densities are $S^{a}(\bbox{x})={1\over 2}\rho (\bbox{x})s^{a}(\bbox{x})$.  The Zeeman term is 
$H_{Z} = -g^{*}\mu _{B}B \int d^{2}x S^{z}(\bbox{x})$.  The ground state $|g_{0}\rangle $ is unique, upon which 
$n^{\upA }=1$ and $n^{\dnA }=0$.  Each Landau level contains two energy levels with the 
one-particle gap energy $g^{*}\mu _{B}B$.  

When the wave function is factorized as in (\ref{SkyrmWave}), the one-point 
function is given by $\langle \varphi ^{\alpha }(\bbox{x})\rangle \equiv \omega ^{\alpha }(z)$.  Based on the formula (\ref{BasicFormu}) it 
is parametrized as
\begin{equation}
e^{-\a(\bbox{x})}e^{i\chi (\bbox{x})} \sqrt {\rho (\bbox{x})}n^{\alpha }(\bbox{x}) = \omega ^{\alpha }(z) ,
\label{ClassGenerB}
\end{equation}
from which the soliton equation (\ref{SolitEq}) follows as the Caucy-Rieman 
equation \cite{EzaIQC}.

On the other hand, the effective Hamiltonian describing the 
perturbative property of the spin texture is derived as follows \cite{EzaIQC,MG}.  
We make the LLL projection of the spin texture, which results in 
$|\FF\rangle =e^{i\Q}|g_{0}\rangle $ where $\Q$ is a LLL projected SU(2) generator.  We evaluate 
the energy of the state $|\FF\rangle $ and identify it with the effective 
Hamiltonian, $H_{\text{eff}}=\langle \FF|(H_{C}+H_{Z})|\FF\rangle $, where $H_{C}$ is the Coulomb 
interaction term.  Making a perturbative expansion in terms of the sigma 
field, making a gradient expansion and taking only the lowest order terms, we 
obtain\cite{EzaIQC,MG}
\begin{equation}
H_{\text{eff}}= {1\over 2}\rho _{s}\int \dx\biggl\{[\partial _{k}\bbox{s}(\bbox{x})]^{2} - {\rho _{0}\over \rho _{s}}g^{*}\mu _{B}B s^{z}(\bbox{x})\biggr\},
\label{EnergChangSPN}
\end{equation}
as yields the NL$\sigma $ model\cite{SkyrmQH}.  The first term represents the spin 
stiffness \cite{KallinHalperin} with $\rho _{s}=\nu e^{2}/(16\sqrt {2\pi }\varepsilon \ell _{B})$. 

We consider the vanishing limit of the Zeeman term ($g^{*}=0$).  In this 
case the ground state is given by an arbitrary constant sigma field, 
$\bbox{s}(\bbox{x})$=constant.  All spins are polarized into one arbitrary direction.  There 
exists a degeneracy in the ground states.  The choice of a ground state 
implies a spontaneous magnetization, or a QH ferromagnetism.  When a 
continuous symmetry is spontaneously broken, there arises a gapless mode known 
as the Goldstone mode.  Quantum coherence develops spontaneously.  Actually, 
due to the Zeeman term all spins are polarized into the $z$ axis.  As far as 
the Zeeman effect is small enough, the system is still considered as a QH 
ferromagnet with a finite coherent length.  The Goldstone mode describes small 
fluctuations of the CP field around the ground state $|g_{0}\rangle $.  We may set 
$n^{\upA }(\bbox{x})=1$ and $n^{\dnA }(\bbox{x})={\zeta (\bbox{x})/\sqrt {\rho _{0}}}$, with $[\zeta ^{\dagger }(\bbox{x}),\zeta (\bbox{y})]=i\delta (\bbox{x}-\bbox{y})$.  The 
effective Hamiltonian (\ref{EnergChangSPN}) implies\cite{EzaIQC}
\begin{equation}
H^{\zeta }_{\text{eff}}=  {2\rho _{s}\over \rho _{0}}\int d^{2}x[\partial _{k}\zeta ^{\dagger }(\bbox{x})\partial _{k}\zeta (\bbox{x}) + \xi _{L}^{-2} \zeta ^{\dagger }(\bbox{x})\zeta (\bbox{x})] ,
\label{EffecGold}
\end{equation}
where $\xi _{L}=\sqrt {2\rho _{s}/g^{*}\mu _{B}B\rho _{0}}$ is the coherent length.
{\par{\bf Activation Energy}:\ }
We analyze the soliton equation (\ref{SolitEq}) for the skyrmion excitation with 
the topological density (\ref{SolitEqSkyrmB}).  Approximate solutions are 
constructed in the two limits, the large skyrmion limit ($\kappa \gg 1$) and the small 
skyrmion limit ($\kappa \ll 1$).  First, in the large limit we can solve (\ref{SolitEq}) 
iteratively, where the first order term is
\begin{equation}
\varrho  (\bbox{x}) = \rho (\bbox{x}) - \rho _{0} \simeq  -\nu Q^{P}(\bbox{x}),
\label{LargeSkyrmSolut}
\end{equation}
with the Pontryagin density (\ref{SolitEqSkyrmB}).  This agrees with the formula 
due to Sondhi et al.\cite{SkyrmQH}.  However, in the small limit the topological 
charge $Q^{P}(\bbox{x})$ is localized within the core:  Indeed, we have $Q^{P}(\bbox{x})\rightarrow \delta (\bbox{x})$ as 
$\kappa \rightarrow 0$ in (\ref{SolitEqSkyrmB}), with which the skyrmion is reduced to the vortex.

We evaluate the excitation energy of one skyrmion.  In the 
semiclassical approximation it consists of the electrostatic energy $E_{C}$ and 
the Zeeman energy $E_{Z}$,
\begin{equation}
E_{\text{skyrmion}} = E_{C} + {1\over 2}g^{*}\mu _{B}B \Delta N_{s} ,
\label{SkyrmEnergSpin}
\end{equation}
where
\begin{equation}
E_{C} = {e^{2}\over 2\varepsilon }\int \dx\dy {\varrho  (\bbox{x})\varrho  (\bbox{y})\over |\bbox{x}-\bbox{y}|},
\label{CouloEnerg}
\end{equation}
and the skyrmion spin $\Delta N_{s}$ is
\begin{equation}
\Delta N_{s}= \int \dx \bigl\{\rho _{0}-\rho (\bbox{x})s^{z}(\bbox{x})\bigr\} .
\label{NumbeFlip}
\end{equation}
In evaluating the Coulomb energy we use (\ref{LargeSkyrmSolut}) for a large 
skyrmion to obtain $E_{C}=\nu ^{2}(\beta /\kappa )E_{C}^{0}$ with $\beta =3\pi ^{2}/64$.  For a small skyrmion we 
solve numerically the soliton equation (\ref{SolitEq}) with $Q(\bbox{x})=\delta (\bbox{x})$, and obtain 
$E_{C}=\nu ^{2}\gamma E_{C}^{0}$ with $\gamma \simeq 0.39$.  

The skyrmion spin $\Delta N_{s}$ would diverge logarithmically for the above 
skyrmion \cite{SkyrmQH}.  This is a fake since the the Zeeman term breaks the spin 
SU(2) symmetry explicitly and introduces a coherent length $\xi _{L}$ into the SU(2) 
component.  The skyrmion configuration is valid only within the coherent 
domain because the coherent behavior of the spin texture is lost outside it.  
By cutting the upper limit of the integration at $r\simeq \xi _{L}$ in (\ref{NumbeFlip}), we 
obtain
\begin{equation}
\Delta N_{s} = \kappa ^{2} \ln\biggl({\xi _{L}^{2}\over \kappa ^{2}\ell _{B}^{2}}+1\biggr) ,
\label{NumbeSpinFlip}
\end{equation}
with the coherent length $\xi _{L}$ given below (\ref{EffecGold}).

We consider the integer QH state at $\nu =1$.  The energy of the skyrmion 
is explicitly given by
\begin{equation}
E_{\text{skyrmion}} = \biggl\{{\beta \over \kappa } + {\widetilde{g}\kappa ^{2}\over 2}\ln\biggl({\sqrt {2\pi }\over 8\widetilde{g}\kappa ^{2}}+1\biggr)\biggr\} E_{C}^{0} ,
\label{SkyrmTotalEnerg}
\end{equation}
where $\widetilde{g}=g^{*}\mu _{B}B/E_{C}^{0}$ is the Zeeman energy in unit of the Coulomb energy $E_{C}^{0}$.  
By minimizing this energy the skyrmion scale $\kappa $ is determined as 
\begin{equation}
\kappa  \simeq  \beta ^{1/3}\biggl\{\widetilde{g}\ln\biggl({\sqrt {2\pi }\over 32\widetilde{g}}+1\biggr)\biggr\}^{-1/3} ,
\label{OptimScale}
\end{equation}
Our main result is the energy formula (\ref{SkyrmTotalEnerg}) with (\ref{OptimScale}).  
It is notable that the skyrmion energy vanishes as $\widetilde{g}\rightarrow 0$, where the skyrmion 
scale is infinitely large.  

The activation energy of a skyrmion-antiskyrmion pair will be given by 
$2E_{\text{skyrmion}}$, which should be compared with experimental data.  Our 
theoretical estimation is about two times bigger than the observed data for 
sample SI1 due to Schmeller et al.\cite{SkyExpEneA}.  There are two main reasons 
for the discrepancy:  First, we have neglected a finite thickness of the 
layer.  Second, we have assumed that the Coulomb energy of the antiskyrmion is 
the same as the skyrmion.  Hence, the value $\beta $ representing the strength of 
the Coulomb energy is subject to a modification in (\ref{SkyrmTotalEnerg}).  We 
have fitted the date by adjusting the parameter $\beta $ in Fig.\ref{SkyrEneTPS}.  
Our theoretical curve reproduces the data remarkably well.  There is a small 
deviation for $\widetilde{g}>0.03$, where the large-skyrmion approximation is no longer 
valid.  The skyrmion spin is estimated that $\Delta N_{s}\simeq 8.7$ at $\widetilde{g}=0.01$ and $\Delta N_{s}\simeq 
5.6$ at $\widetilde{g}=0.02$.

{\par{\bf Discussions}:\ }
We comment on the spin-stiffness term in (\ref{EnergChangSPN}) since our formula 
(\ref{SkyrmEnergSpin}) is different from the standard one \cite{SkyrmQH,SkyrmA} by this 
term.  The term has been derived perturbatively from the microscopic 
Hamiltonian \cite{EzaIQC,MG,KallinHalperin}.  It describes correctly perturbative 
long-distance physics.  However, there is no reason that it describes a 
nonperturbative object such as skyrmions.  As we have argued, its absence is 
required from the consistency condition that the skyrmion wave function is 
reduced to the vortex wave function in the limit $\kappa \rightarrow 0$.  Indeed, the skyrmion 
energy is reduced to the vortex energy without this term.  We also emphasize 
that the experimental data are excellently fitted by the skyrmion energy 
(\ref{SkyrmEnergSpin}) without it.  A decisive test is the vanishing of the 
activation energy at $\widetilde{g}=0$ in pure samples, which seems to be confirmed in a 
recent experiment \cite{SkyExpEneB}.

\begin{figure}[thb]
\caption{
Our theoretical curves versus experimental data due to Schmeller et al.[5].  
The data for the sample SI1, which has a high mobility, are fitted excellently 
by our formula (27) with the choice of $\beta =0.172$.  The same theoretical curve 
fits also the data for other samples with low mobilities excellently by 
introducing appropriate offsets $\Gamma $ to take care of the level broadening.  The 
activation energy $\Delta $ determined by the Arrhenius plots includes in general a 
sample dependent offset that increases with disorder [6], 
$E_{\text{skyrmion-pair}}=\Delta +\Gamma $.}
\label{SkyrEneTPS}
\end{figure}

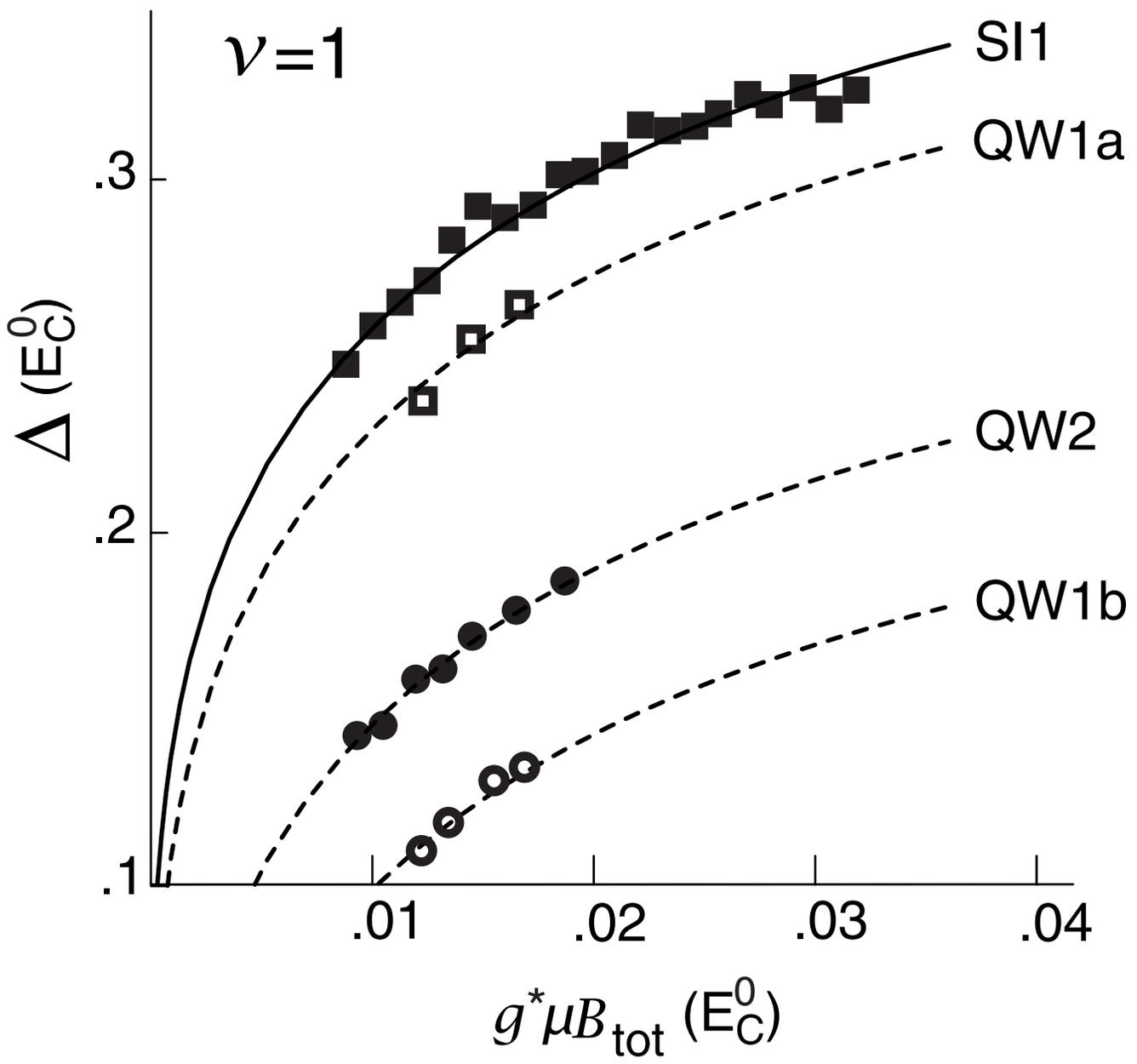
\end{document}